\documentstyle[12pt]{article}

\begin{document}
\rightline{ILL-(TH)-94-17}
\bigskip\bigskip
\begin{center}
{\Large \bf Quark-Lepton Unification and Rare Meson Decays}\\
\bigskip\bigskip\bigskip\bigskip
{\large \bf G. Valencia}\\
\medskip
Department of Physics\\Iowa State University\\Ames, IA   50011\\
\bigskip\bigskip
{\large \bf S. Willenbrock}\\
\medskip
Department of Physics\\University of Illinois\\1110 West Green Street\\
Urbana, IL   61801
\end{center}
\bigskip\bigskip\bigskip\bigskip

\begin{abstract}
We study meson decays mediated by the heavy gauge bosons of the
Pati-Salam model of quark-lepton unification.  We consider the scenarios
in which the $\tau$ lepton is associated with the third, second, and first
generation of quarks.  The most sensitive probes, depending on the
scenario, are rare $K$, $\pi$, and $B$ decays.
\end{abstract}

\newpage

\addtolength{\baselineskip}{9pt}

\section{Introduction}

\indent\indent One of the unexplained features of the standard model of
the strong and electroweak
interactions is why some fermions, the quarks, experience the strong
interaction
while others, the leptons, do not. Experience has taught us to look for
symmetry even
when it is not apparent, and this leads one to speculate that, at some deeper
level, quarks and leptons are identical.  Perhaps there exists a symmetry
between
quarks and leptons which is broken at high energy, in much the same way that
the
electroweak symmetry is broken at an energy scale of
$(\sqrt{2} G_{F})^{-1/2} \approx$ 250 GeV.

If we further speculate that the quark-lepton symmetry is a local gauge
symmetry,
we are led to predict a new force of nature which mediates transitions between
leptons and quarks.  The simplest model which incorporates this idea is the
Pati-Salam model \cite{Pati74}, based on the group $SU(4)_{c}$.  The subgroup
$SU(3)_{c}$ is the ordinary strong interaction, and lepton number is the fourth
``color''.  At some high energy scale, the group $SU(4)_{c}$ is
spontaneously broken
to $SU(3)_{c}$, liberating the leptons from the influence of the strong
interaction and breaking the symmetry between quarks and leptons.

In this paper we explore signals for quark-lepton unification \`{a} la
Pati-Salam.  We show that rare $K$, $\pi$, and $B$
decays are the most sensitive probes of the presence of quark-lepton
transitions mediated by heavy Pati-Salam bosons.  A new feature of our
analysis is
that we do not restrict ourselves to the assumption that the $\tau$ lepton is
associated with the third generation of quarks, but also consider the
possibility that it is associated with the second, or even first,
generation. A recent paper on Pati-Salam bosons also considers
this possibility \cite{KM}.  Our analyses overlap for $K_L \to
\mu^{\pm}e^{\mp}$
and $\Gamma(\pi^+ \to e^+\nu)/\Gamma(\pi^+ \to \mu^+\nu)$, and agree.  We
further show that $\Gamma(K^+ \to e^+\nu)/\Gamma(K^+ \to \mu^+\nu)$ and rare
$B$ decays are the most sensitive probes in the scenario in which the $\tau$
lepton is associated with the first generation of quarks.

Pati-Salam bosons are members of a class of bosons called ``leptoquarks'',
since they mediate transitions between leptons and quarks.  They are spin
one, and have non-chiral couplings to quarks and leptons.  There are several
recent model-independent analyses of bounds on leptoquarks.
Ref.~\cite{Leurer93} concentrates on spin-zero leptoquarks with chiral
couplings, and Ref.~\cite{Leurer94}
on spin-one leptoquarks with chiral couplings. Ref.~\cite{Davidson94} considers
both spin-zero and spin-one leptoquarks, with chiral and non-chiral couplings,
and with the leptons associated with the quark generations in all six
permutations.

In Section 2 we review the Pati-Salam model.  In Sections 3, 4, and 5 we
discuss
rare decays mediated by Pati-Salam bosons in the scenarios where the $\tau$
lepton is associated with the third, second, and first generations,
respectively. The bounds on the Pati-Salam-boson mass from rare $K$, $\pi$, and
$B$ decays are summarized in Table 1. Section 6 contains our conclusions.

\section{Pati-Salam Model}

\indent\indent Pati and Salam proposed a class of unified models which
incorporate quark-lepton unification \cite{Pati74}.$^{}$\footnote{For a review,
see Ref.~\cite{Langacker81}.}  A common feature of these models is the group
$SU(4)_{c}$, with the subgroup $SU(3)_{c}$ corresponding to the strong
interaction, and with lepton
number identified as the fourth ``color''.  In this section we discuss the
minimal model which embodies quark-lepton unification via the $SU(4)_{c}$
Pati-Salam group.

Because quarks and leptons with the same $SU(2)_L$ quantum number have
different hypercharge, the Pati-Salam group $SU(4)_{c}$ cannot commute with
hypercharge.  Furthermore, although $SU(4)_{c}$ can break to
$SU(3)_{c}\times U(1)$, this $U(1)$ is not hypercharge, but rather the
difference of baryon number and lepton number; we henceforth refer to it as
$U(1)_{B-L}$, as
is standard.  Another group is needed to replace hypercharge.  The simplest
possibility is to introduce another $U(1)$ group, called $U(1)_{T_{3R}}$
(notation to be explained shortly), such that
$U(1)_{B-L} \times U(1)_{T_{3R}}$ breaks spontaneously to $U(1)_Y$.

The particle content of the $SU(4)_{c} \times SU(2)_{L} \times
U(1)_{T_{3R}}$ model is
\begin{eqnarray}
\left(\begin{array}{clcr}
u_{R} & u_{G} & u_{B} & \nu \\
d_{R} & d_{G} &d_{B} &e\end{array} \right)_{L} & \quad (4, 2, 0) \nonumber \\
\left(\begin{array}{clcr}
u_{R} & u_{G} & u_{B} & \nu\end{array} \right)_{R} & \qquad (4, 1, +
\frac{1}{2})  \nonumber\\
\left(\begin{array}{clcr}
d_{R} & d_{G} & d_{B} & e\end{array} \right)_{R}  & \qquad (4, 1, -
\frac{1}{2}) \nonumber
\end{eqnarray}
where the subscripts on the quarks denote color (red, green,
blue), and
the subscripts $L, R$ denote chirality.  The model is free of gauge and
mixed gravitational anomalies.  The $U(1)_{T_{3R}}$ quantum numbers of the
$SU(2)_L$ singlet fields, $\pm\frac{1}{2}$, suggest that $U(1)_{T_{3R}}$
is a subgroup of an $SU(2)_R$ group; hence the notation.  We
will not make this additional assumption, since it does not affect our
analysis.  However, we remark that
$SU(4)_c\times SU(2)_L \times SU(2)_R$ is a maximal subgroup of
$SO(10)$, so another motivation for considering $SU(4)_c$ is $SO(10)$ grand
unification \cite{Georgi74,Fritzsch75}.  However, the $SU(4)_c$ breaking
scale in this model is very high, at least $10^{11}$ GeV,
well out of reach of low-energy experiments \cite{Deshpande92}.

Another motivation for considering the Pati-Salam group is provided by
extended technicolor models.  One can show that these models must
incorporate gauged quark-lepton unification, or massless neutral
Goldstone bosons (axions) and light ($\sim$ 5 GeV) charged
pseudo-Goldstone bosons will result from electroweak symmetry
breaking \cite{Eichten80}.  The simplest way to achieve this, often employed
in model building \cite{Dimopoulos80,Lykken94,Appelquist93},
is to introduce a Pati-Salam group.

One canonically associates the $\tau$ lepton with the third generation of
quarks, both
for reasons of mass (they are the heaviest known fermions of their respective
classes), and for historical reasons (the $\tau$ lepton and the $b$ quark were
the
last fundamental fermions discovered; evidence for the top quark has
recently been presented \cite{CDF}). This is certainly a natural assumption.
However, the flavor-symmetry-breaking
mechanism, which is responsible for fermion mass generation, is a mystery.  One
should keep an open mind to the possibility that the $\tau$ lepton is actually
associated with the second or first generation of quarks.

Generically, one would expect that there is a mixing matrix, analogous to the
Cabibbo-Kobayashi-Maskawa (CKM) matrix, which describes the mixing of the
lepton
generations with the quark generations.  We will make the assumption that this
matrix is nearly diagonal, as is the CKM matrix, but consider the scenarios
where the $\tau$ lepton is most closely associated with the third, second or
first
generations in the following sections.

Because the Pati-Salam interaction conserves $B-L$ and fermion number, it
cannot
mediate nucleon decay.  Purely leptonic transitions, such as $\mu \to
e \gamma$ and $\mu N\to eN$, and meson-antimeson mixing, are induced
only at one loop, and vanish in the limit of zero intergenerational mixing.
The natural place to search for the Pati-Salam interaction is therefore in
meson decays.  These will be considered in the following sections.

\section{Tau Lepton Associated With Third-Generation Quarks}

\indent\indent The long-lived kaon, due to its longevity, is a
sensitive probe of suppressed interactions which produce unusual decays.
Pati and Salam observed that the decay $K_L\to \mu^{\pm}e^{\mp}$,
shown in Fig.~1, provides the best bound on the mass of the
Pati-Salam bosons \cite{Pati74}. This
bound was later refined in Refs.~\cite{Dimopoulos93,Shanker206},
and leading-log QCD effects were included in Ref.~\cite{Deshpande83}.
Here we update this bound, based on the recent
upper bound BR ($K_{L} \to \mu^{\pm} e^{\mp}) < 3.9\times 10^{-11}$
(90$\%$ C.L.) from Brookhaven E791 \cite{Arisaka93}.  Combined with previous
experiments, this yields
\begin{equation}
BR(K_{L} \to \mu^{\pm} e^{\mp}) < 3.3 \times 10^{-11} \quad(90\% \;
\rm{C.L.})\;.
\label{BNL}
\end{equation}

The diagram in Fig.~1 gives rise to an effective four-fermion interaction\\
\begin{equation}
{\cal L}_{eff} = \frac{g_4^2}{2M_c^2} \overline{d} \gamma^{\mu} e
\overline{\mu}
\gamma_{\mu} s + h.c.
\end{equation}
where a sum on color is implicit.  $M_c$ is the mass of the
Pati-Salam bosons, and $g_4$ is the Pati-Salam coupling at the scale $M_c$.
Since $SU(4)_c$ breaks to $SU(3)_c$, this coupling is equal to the strong
coupling at $M_c$. A Fierz rearrangement gives
\begin{equation}
{\cal L}_{eff} = \frac{g_4^2}{2M_c^2} \left[- \overline{d} s\overline{\mu} e
+ \frac{1}{2} \overline{d} \gamma^{\mu} s\overline{\mu} \gamma_{\mu} e
+ \frac{1}{2} \overline{d} \gamma^{\mu} \gamma_5 s\overline{\mu}\gamma_{\mu}
\gamma_5 e + \overline{d} \gamma_5 s\overline{\mu} \gamma_5 e + h.c.\right]\;.
\label{EFF}
\end{equation}

For $K_L\to \mu^{\pm}e^{\mp}$, the required matrix
elements are
\begin{eqnarray}
<0 | \overline{d} \gamma^{\mu} \gamma_5 s | \overline{K}^0 (p) > & = &
i\sqrt{2} F_K p^{\mu} \quad(F_K = 114 \; \rm{MeV}) \\
<0 | \overline{d} \gamma_5 s | \overline{K}^0 (p) > & = & -i \sqrt{2}B_0 F_K
\end{eqnarray}
where \cite{Donoghue92}
\begin{equation}
B_0 = \frac{m_K^2}{m_s +m_d}
\end{equation}
and $m_s, m_d$ are the running $\overline{MS}$ quark masses evaluated at
the Pati-Salam scale.  These masses are evolved to low energy using
leading-log QCD evolution \cite{Deshpande83},
\begin{equation}
m(\mu)=m(M_c)\left(\frac{\alpha_s(m_t)}{\alpha_s(M_c)}\right)^{4/7}
\left(\frac{\alpha_s(m_b)}{\alpha_s(m_t)}\right)^{12/27}
\left(\frac{\alpha_s(m_c)}{\alpha_s(m_b)}\right)^{12/25}
\left(\frac{\alpha_s(\mu)}{\alpha_s(m_c)}\right)^{4/9} \;.
\end{equation}

The light-quark $\overline{MS}$ masses are not well known, although their
ratios are known
from chiral perturbation theory: $m_u/m_d = 0.56$, $m_s/m_d = 20.1$, at
leading order \cite{Weinberg77}.  The absolute scale of the quark masses must
be
obtained from nonperturbative QCD.  Lattice gauge theory provides a rough
estimate of the light-quark $\overline{MS}$ masses.\footnote{Although we expect
to eventually know the value of the light-quark $\overline{MS}$ masses from
lattice calculations, at present these masses are not known with any accuracy.}
{}From Ref.~\cite{Ukawa92}, we estimate
$\hat m = (m_u + m_d)/2 = 2-5$ MeV at $\mu =$ 1 GeV. Since
the quark masses enter in the denominator of $B_0$, we conservatively use
the high values: $m_d =$ 7.7 MeV, $m_s =$ 125 MeV, at $\mu =$ 1 GeV.

The partial width for $K_L \to \mu^{\pm} e^{\mp}$ is\footnote{The notation
indicates a sum over the $\mu^+e^-$ and $\mu^-e^+$ final states.}
\begin{equation}
\Gamma (K_L \to \mu^{\pm} e^{\mp}) = \pi\alpha_s^2
(M_c) \frac{1}{M_c^4} F_K^2 m_K B_0^2
\left(1 - \frac{m_{\mu}^2}{m_K^2}\right)^2 \;.
\end{equation}
We use $\alpha_s(M_Z)=$.115 ($\Lambda_4 =$ 0.275 MeV), evolved to $M_c$ via the
two-loop renormalization group (with $m_t=$170 GeV), assuming no other colored
particles lie between $m_t$ and $M_c$ (such particles would increase
$\alpha_s(M_c)$ and increase the lower bound on $M_c$).  Using the upper bound
on $K_L\to \mu^{\mp}e^{\pm}$ of Eq.~(\ref{BNL}) we find
\begin{equation}
M_c>1400 \; \rm{TeV}.
\end{equation}
It is remarkable that physics at such a high scale can be probed
by this decay.  Future experiments may probe branching ratios as small as
$10^{-14}$, increasing the lower bound on $M_c$ by a factor of about 7.

The Pati-Salam bosons also produce transitions between
bottom quarks and $\tau$ leptons.  If we replace the $s$ quark and muon in
Fig.~1 with a
$b$ quark and $\tau$ lepton, we obtain the diagram for the decay
$\overline{B}_d^0
\to \tau^- e^+$.   The partial width is
\begin{equation}
\Gamma(\overline{B}_d^{0} \to \tau^- e^+) = \pi \alpha_s^2 (M_c)
\frac{1}{M_c^4} F_B^2 m_B^3 \left(R - \frac{1}{2} \frac{m_{\tau}}{m_B}\right)^2
\left(1 - \frac{m_{\tau}^2}{m_B^2}\right)^2
\label{BTAUE}
\end{equation}
where
\begin{equation}
R=\frac{m_B}{m_b}
\left(\frac{\alpha_s(M_c)}{\alpha_s(m_t)}\right)^{4/7}
\left(\frac{\alpha_s(m_t)}{\alpha_s(m_b)}\right)^{12/27} \label{R}
\end{equation}
and $m_b$ is the $\overline{MS}$ mass evaluated at $\mu = m_b$.  This is
known from lattice-QCD calculations of the $\Upsilon$ spectrum to be about
$m_b(m_b)= 4.3$ GeV.

The experimental upper bound on this decay from CLEO is \cite{CLEO93}
\begin{equation}
BR(\overline{B}_d^0 \to \tau^{\pm} e^{\mp}) < 5.3 \times 10^{-4}
\quad(90\% \;\rm{C.L.})\;.
\end{equation}
Using $F_B =$ 140 MeV and $\tau_{B^0} = 1.3 \;ps$ we find
\begin{equation}
M_c > 4.8 \;{\rm TeV} \; ,
\end{equation}
much less than the lower bound on $M_c$ from
$K_L \to \mu^{\pm} e^{\mp}$.

We have also considered all other meson decays mediated by Pati-Salam bosons:
$\pi^+ \to e^+\nu$;
$\pi^0 \to e^+ e^-, \nu \overline{\nu}$; $K^+ \to \mu^+ \nu$;
$D^+ \to e^+ \nu$; $D^0\to\nu\overline{\nu}$;
$D_s^+\to \mu^+\nu$; $B^+\to \tau^+\nu$;
$B_c \to \tau^+ \nu$; and $\overline{B}_s^0\to\tau^-\mu^+$.
None competes with $K_{L} \to
\mu^{\pm} e^{\mp}$ in its sensitivity to the Pati-Salam interaction.

If we associate the muon with the first generation of quarks and the
electron with
the second, the relevant decays are $K_L\to\mu^{\pm}e^{\mp}$;
$\overline{B}_d^0\to\tau^-\mu^+$; etc.  The best bound on $M_c$ again
comes from $K_L\to\mu^{\pm}e^{\mp}$.

\section{Tau Lepton Associated with Second-Generation Quarks}

\indent\indent At first sight, associating the $\tau$ lepton with the second
generation of quarks and, say, the muon with the third generation seems
unnatural. However, the $\tau$ lepton is comparable in mass to the
second-generation charm quark.  Although the muon is a factor of about 40
less massive than the bottom quark,
the bottom quark is at least a factor of 30 less massive than the top quark
($m_t > 131$ GeV \cite{D0}), so large intragenerational mass ratios
do occur.

Because the strange quark is associated with the $\tau$ lepton, the decay of
$K_L$ to leptons does not occur via the Pati-Salam interaction.
Pati-Salam bosons also mediate transitions between up quarks and
neutrinos, so if we replace the $s$ quark and muon in Fig. 1 with an up quark
and electron neutrino, we obtain the diagram for
$\pi^{+} \to e^{+} \nu_{e}$.  This process involves only first-generation
quarks. Since the decay
$\pi^{+} \to e^{+} \nu_{e}$ also proceeds via the weak interaction, the
presence of a contribution from the Pati-Salam interaction manifests itself as
a
violation of lepton universality in $R_{e/\mu} = \Gamma
(\pi^{+}\to e^{+}\nu)/\Gamma(\pi^+\to\mu^+\nu)$
\cite{Shanker204}.  The
theoretical prediction from the weak interaction is \cite{Marciano93}
\begin{equation}
R_{e/\mu}^{theory} = (1.2352 \pm .0005) \times 10^{-4}
\end{equation}
while the current experimental measurements are
\begin{eqnarray}
R_{e/\mu} & = & (1.2265 \pm .0034 \pm .0044) \times 10^{-4} \quad
({\rm TRIUMF}\; \cite{Britton92}) \\
R_{e/\mu} & = & (1.2346 \pm .0035 \pm .0036) \times 10^{-4} \quad
({\rm PSI}\; \cite{Czapek93})
\end{eqnarray}
which combined gives
\begin{eqnarray}
R_{e/\mu} & = & (1.2310 \pm .0037) \times 10^{-4} \;.
\end{eqnarray}
The theoretical uncertainty is much less than the experimental
uncertainty.

The contribution of the Pati-Salam interaction to $\pi^{+} \to
e^{+} \nu_{e}$
is obtained via the interference of the Pati-Salam and weak amplitudes.  We
find
\begin{equation}
\Delta \Gamma (\pi^+ \to e^+ \nu_e) = - \alpha_s (M_c)
\frac{G_F}{\sqrt{2}} F_{\pi}^2 \frac{m_e m_{\pi}}{M_c^2} V_{ud} B_0
\label{DELTA}
\end{equation}
to be compared with the tree-level weak decay width
\begin{equation}
\Gamma (\pi^{+} \to e^{+} \nu_{e}) = \frac{1}{4\pi} G_F^{2} F^{2}_{\pi}
m^{2}_{e} m_{\pi} \left| V_{ud} \right|^{2}\; .
\end{equation}
The absence of a deviation of the theoretical prediction from the
experimental measurements yields a lower bound on the mass of the
Pati-Salam boson of
\begin{equation}
M_c > 250 \;{\rm TeV}\; . \label{PI}
\end{equation}
This bound is a factor of about five less stringent than the
bound from
$K_{L} \to \mu^{\pm}e^{\mp}$ in the previous section.  Nevertheless,
it is the strongest bound for the scenario considered here.

If we replace the $s$ quark in Fig. 1 with a $b$ quark, we obtain the
diagram for $\overline{B}_{d}^{0} \to \mu^{-}e^{+}$.  The partial
width is obtained from Eq.~(\ref{BTAUE}),
\begin{equation}
\Gamma (\overline{B}^0_d \to \mu^- e^+) = \pi
\alpha^2_s (M_c) \frac{1}{M_c^4}F_B^2 m_B^3 R^2 \label{BMUE}
\end{equation}
where we have neglected the lepton masses. The present upper
bound on this decay from CLEO \cite{CLEO93}
\begin{equation}
BR (\overline{B}^0_d \to \mu^{\pm}e^{\mp}) < 5.9 \times 10^{-6}
\end{equation}
places a lower bound on the Pati-Salam-boson mass of
\begin{equation}
M_c > 16 \;{\rm TeV}\;.
\label{BD10}
\end{equation}
The upper bound on this decay can be
significantly improved with $B^0_d$ mesons produced in hadron colliders.
A lower bound on the branching ratio of $10^{-9}$ translates into $M_c >
140$ TeV.

If we further replace the $d$ antiquark in Fig. 1 with a $u$ antiquark and the
positron with an antineutrino, we obtain the (charge conjugate of the)
diagram for $B^+ \to \mu^+\nu_e$.  The partial
width is the same as Eq.~(\ref{BMUE}):
\begin{equation}
\Gamma (B^+ \to \mu^+\nu) = \pi
\alpha^2_s (M_c) \frac{1}{M_c^4}F_B^2 m_B^3 R^2 \;.\label{BMUNU}
\end{equation}
The present upper
bound on this decay from CLEO \cite{Cinabro92}
\begin{equation}
BR (B^+ \to \mu^+\nu) < 2.0 \times 10^{-5}
\end{equation}
places a lower bound on the Pati-Salam-boson mass of
\begin{equation}
M_c > 12 \;{\rm TeV}\;,
\label{BPLUS}
\end{equation}
comparable to the bound from $\overline{B}^0_d \to \mu^-e^+$.

If we replace the muon in Fig. 1 with a $\tau$ lepton, the diagram no longer
describes
$K_L$ decay, but rather $\tau^- \to ``K" e^-$, where $``K"$ denotes
a meson or
mesons with strangeness $-1$.  The effective interaction is the same as
Eq.~(\ref{EFF}), but with the muon replaced by the $\tau$ lepton.  The decay to
the
ground
state, $\tau^- \to K_se^-$, is
\begin{equation}
\Gamma(\tau^- \to K_se^-) = \frac{\pi}{4} \alpha^2_s (M_c)
\frac{1}{M^4_c}
F^2_Km_{\tau} \left(B_0 - \frac{1}{2}m_{\tau}\right)^2
\left(1 - \frac{m^2_K}{m^2_{\tau}}\right)^2 \;.
\label{KSHORT}
\end{equation}
The decay to the first excited state,
$\tau^-
\to \overline{K}^{*0}e^-$, involves only the vector current.  Using
\begin{equation}
< 0\mid \overline{d} \gamma^{\mu}s\mid \overline{K}^{*0} (p) > =
ig_{K^*} \epsilon^{\mu} (p)\;\;\;\; (g_{K^*} = .133 \;{\rm GeV}^2)
\end{equation}
we find
\begin{equation}
\Gamma(\tau^- \to \overline{K}^{*0}e^-) =
\frac{\pi}{8}\alpha^2_s (M_c)\frac{1}{M^4_c}g^2_{K^*}m_{\tau}\left(
1+\frac{1}{2} \frac{m^2_{\tau}}{m^2_{K^*}}\right) \left(
1-\frac{m^2_{K^*}}{m^2_{\tau}} \right)^2 \;.
\label{KSTAR}
\end{equation}
The two decay modes are of comparable sensitivity to Pati-Salam bosons.
The upper bound on $\tau^- \to \overline{K}^{*o} e^-$ from CLEO
\cite{Bartelt94},
\begin{equation}
BR(\tau^- \to \overline{K}^{*0} e^-) < 1.1 \times 10^{-5}
\end{equation}
gives the best lower bound
from $\tau$ decays on the Pati-Salam-boson mass. We find
\begin{equation}
M_c > 1.6 \;{\rm TeV} \; , \label{TAU}
\end{equation}
not nearly as strong as the lower bound from other decays.

If we associate the muon with the first generation of quarks and the
electron with the third, the relevant decays are $\pi^+ \to \mu^+\nu_{\mu}$,
$\overline{B}^0_d \to \mu^+ e^-$, $B^+ \to e^+\nu$,
$\tau^- \to K_s \mu^- $, $\tau^- \to \overline{K}^{*0}\mu^-$,
and $\overline{B}^0_s \to \tau^+ e^-$.  The best lower bound on the
mass of the Pati-Salam bosons again comes from $R_{e/\mu} = \Gamma
(\pi^{+}\to e^{+}\nu)/\Gamma(\pi^+\to\mu^+\nu)$.  Since
the Pati-Salam interaction contributes to the unsuppressed weak decay
$\pi^+ \to \mu^+\nu$, rather than the suppressed decay $\pi^+ \to e^+\nu$
as in the previous case, the bound is not as strong as before.
The interference of the Pati-Salam and weak amplitudes for $\pi^+ \to
\mu^+\nu$ is given by
\begin{equation}
\Delta \Gamma (\pi^+ \to \mu^+ \nu_{\mu}) = - \alpha_s (M_c)
\frac{G_F}{\sqrt{2}} F_{\pi}^2 \frac{m_{\mu}
m_{\pi}}{M_c^2}V_{ud} \left(B_0-\frac{1}{2}m_{\mu}\right)
\left(1-\frac{m_{\mu}}{m_{\pi}}\right)^2
\end{equation}
to be compared with the tree-level weak decay width
\begin{equation}
\Gamma (\pi^{+} \to \mu^{+} \nu_{\mu}) = \frac{1}{4\pi} G_F^{2}
F^{2}_{\pi} m^{2}_{\mu} m_{\pi} \left| V_{ud} \right|^{2}
\left(1-\frac{m_{\mu}}{m_{\pi}}\right)^2\; .
\end{equation}
The absence of a deviation of the theoretical prediction from the
experimental measurements yields a lower bound on the mass of the
Pati-Salam boson of
\begin{equation}
M_c > 76 \;{\rm TeV} \; .
\end{equation}

The bound from $\overline{B}^0_d \to
\mu^+e^-$ is the same as Eq.~(\ref{BD10}).  This mode will ultimately place
the best lower bound on the mass of the Pati-Salam boson using $B^0_d$ mesons
produced in hadron colliders, as mentioned
above. The bound from $B^+ \to e^+\nu$ is about the same as from $B^+ \to
\mu^+ \nu$, Eq.~(\ref{BPLUS}). The bound on the decay
$\tau^- \to \overline{K}^{*0}\mu^-$ is
similar to that with an electron in the final state \cite{Bartelt94},
\begin{equation}
BR(\tau^- \to \overline{K}^{*0}\mu^-) < 8.7 \times 10^{-6}
\end{equation}
and yields approximately the same lower bound on the Pati-Salam-boson mass,
Eq.~(\ref{TAU}).

\section{Tau lepton associated with first-generation quarks}

\indent\indent In this section we discuss the case where the $\tau$ lepton is
associated with the first generation of quarks.
We first assume the muon is associated with the second generation and the
electron with the third.  One might imagine this scenario being realized by a
``see-saw''-type mechanism for quark and lepton masses.

The best current lower bound on the mass of the Pati-Salam boson comes from
$B^+ \to e^+ \nu$, which has the same partial width as $B^+ \to \mu^+ \nu$,
Eq.~(\ref{BMUNU}).  The upper bound on this branching ratio from CLEO
\cite{Cinabro92}
\begin{equation}
BR (B^+ \to e^+\nu) < 1.3 \times 10^{-5}
\end{equation}
places a lower bound on the Pati-Salam-boson mass of
\begin{equation}
M_c > 13 \;{\rm TeV}\;.
\label{BEPLUS}
\end{equation}

The decay $\overline{B}^0_s \to e^-\mu^+$ also occurs via the Pati-Salam
interaction.  There is currently no bound on this decay, but the large
number of $B_s^0$ mesons produced in hadron collisions can potentially be used
to probe
branching ratios as small as $10^{-9}$.  This translates into $M_c > 140$ TeV.

In this scenario, as well as the scenarios in the preceding section, the
Pati-Salam boson mediates charmless semileptonic $B$ decay.  This process is
very suppressed in the standard model due to the small value of $V_{ub}$.
The Pati-Salam and weak decay amplitudes do not interfere because the neutrinos
are different types.
The ratio of the Pati-Salam and weak partial widths in the spectator model
is
\begin{equation}
\frac{\Gamma_{PS}(b \to ue^-\bar \nu_{\tau})}{\Gamma(b \to ue^-\bar \nu_{e})}
= \frac{2\pi^2 \alpha_s^2}{G_F^2M_c^4 \left| V_{ub} \right|^2} \;.
\end{equation}
Using $ \left| V_{ub} \right| > .002 $ and $M_c > 13$ TeV (from
Eq.~(\ref{BEPLUS})) yields a ratio less than $1\%$, which is too small
to observe.

Now consider the scenario in which the electron is associated with the
second generation, and the muon with the third.  In this case the best
bound on the mass of the Pati-Salam boson comes from its contribution to
$K^+ \to e^+\nu$.  This manifests itself as a violation of lepton
universality in $R_{e/\mu} = \Gamma
(K^{+}\to e^{+}\nu)/\Gamma(K^+\to\mu^+\nu)$.  The
theoretical prediction from the weak interaction is\footnote{This is the
leading-order prediction with no electromagnetic radiative correction.
This correction depends on the manner in which bremsstrahlung photons are
dealt with experimentally \cite{GW}. }
\begin{equation}
R_{e/\mu}^{theory} = 2.57 \times 10^{-5}
\end{equation}
while the experimental measurement is \cite{Heintze76}
\begin{equation}
R_{e/\mu} = (2.45 \pm 0.11) \times 10^{-5} \;.
\end{equation}
Unlike the case of $\pi^+ \to e^+ \nu$, the Pati-Salam and weak amplitudes
for $K^+ \to e^+ \nu$ do not interfere, because the neutrinos are
different types. The partial width for $K^+ \to e^+ \nu_{\tau}$
via the Pati-Salam interaction is
\begin{equation}
\Gamma (K^+ \to e^+\nu_{\tau}) = \pi
\alpha^2_s (M_c) \frac{1}{M_c^4}F_K^2 m_K B_0^2
\end{equation}
to be compared with the tree-level weak decay width
\begin{equation}
\Gamma (K^+ \to e^{+} \nu_{e}) = \frac{1}{4\pi} G_F^{2} F^{2}_K
m^{2}_{e} m_K \left| V_{us} \right|^{2}\; .
\end{equation}
The absence of a deviation of the theoretical prediction from the
experimental measurement yields a lower bound on the mass of the
Pati-Salam boson of
\begin{equation}
M_c > 130 \;{\rm TeV} \; .
\end{equation}
The bound from the decay $\overline{B}^0_s \to e^+\mu^-$, discussed above, can
potentially
approach this bound.  The bound from $B^+ \to \mu^+ \nu$ is the same as
Eq.~(\ref{BPLUS}).

\section{Conclusions}

\indent\indent In this paper we have studied rare meson decays induced by the
heavy gauge bosons of the Pati-Salam model of quark-lepton unification.
We have considered the scenarios in which the leptons are associated with
the quark generations in all six permutations.  The lower bounds obtained
on the mass of the Pati-Salam bosons are given in Table 1.  Bounds from
$K_L \to \mu^{\pm} e^{\mp}$ and lepton universality in charged pions decays
are well known, and we have updated them.  We have shown that in the two
scenarios in which the $\tau$ lepton is associated with the first
generation of quarks, the best bounds come from $B^+ \to e^+ \nu$ and
lepton universality in charged kaon decays.  All of these measurements
have the potential for improvement.

At present, the bounds from $B_d^0,B_s^0 \to \mu^{\pm} e^{\mp}$ are not
the strongest in any of the scenarios.  However, the large number of these
mesons which are produced in hadron colliders can potentially be used to
probe branching ratios as small as $10^{-9}$.  The resulting bound on the
Pati-Salam-boson mass would be the best for three of the scenarios. A
high-resolution silicon vertex detector is essential for such a measurement.

\section*{Acknowledgments}

S.~W. is grateful for conversations with W. Bardeen, E. Braaten, G. Burdman,
S. Errede, A. Falk, G. Gollin, L. Holloway, T. LeCompte, K. Lingel,
P. MacKenzie, W. Marciano, P. Pal, R. Patterson, M. Selen, J. Urheim, and
C. White. G.~V. thanks W. Bardeen, G. Burdman, and W. Marciano for
conversations. The work of G.~V. was supported in part by a DOE OJI award.
S. W. thanks the high-energy theory groups at Ohio State
University and Fermilab for their hospitality.

\newpage

\begin{table}[hbt]
\caption[fake]{Lower bound on the mass of the Pati-Salam boson (TeV) from rare
$K$, $\pi$, and $B$ decays.  The first column indicates how the leptons are
associated with the first, second, and third generation of quarks.
The best bound for each scenario is
enclosed in a box. The bounds assuming $BR(B_d^0,B_s^0 \to
\mu^{\pm}e^{\mp}) < 10^{-9}$ are shown in parentheses. A dash
indicates the decay does not occur via the Pati-Salam interaction.}
\bigskip
\begin{center}
\begin{tabular}{cccccccc}
&$K_L \to \mu^{\pm}e^{\mp}$&$\frac{\pi^+ \to e^+\nu}{\pi^+ \to \mu^+\nu}$
&$\frac{K^+ \to e^+\nu}{K^+ \to \mu^+\nu}$&$B_d^0 \to \mu^{\pm}e^{\mp}$
&$B_s^0 \to \mu^{\pm}e^{\mp}$&$B^+ \to e^+\nu$&$B^+ \to \mu^+\nu$ \\
\\
$e\mu\tau$  & \fbox{1400} & 250        & 4.9       & -  & -   & -  & -  \\
$\mu e\tau$ & \fbox{1400} & 76         & 130       & -  & -   & -  & -  \\
$e\tau\mu$  & -           & \fbox{250} & -         & 16(140) & -   & -  & 12 \\
$\mu\tau e$ & -           & \fbox{76}  & -         & 16(140) & -   & 13 & -  \\
$\tau\mu e$ & -           & -          & 4.9       & -  & (140) & \fbox{13}&
-\\
$\tau e\mu$ & -           & -          & \fbox{130}& -  & (140) & -  & 12 \\
\end{tabular}
\end{center}
\end{table}

\newpage

\section*{Figure Captions}

\begin{description}
\item[] Fig. 1 - Diagram for $\overline{K}^0 \to \mu^-e^+$, mediated
by a heavy Pati-Salam boson.
\end{description}


\newpage

\begin{thebibliography}{99}

\bibitem{Pati74} J. Pati and A. Salam, Phys. Rev. D {\bf 10}, 275 (1974).

\bibitem{KM} A. Kuznetsov and N. Mikheev, Phys. Lett. {\bf B329}, 295 (1994).

\bibitem{Leurer93} M. Leurer, Phys. Rev. Lett. {\bf 71}, 1324 (1993);
Phys. Rev. D {\bf 49}, 333 (1994).

\bibitem{Leurer94} M. Leurer, Phys. Rev. D {\bf 50}, 536 (1994).

\bibitem{Davidson94} S. Davidson, D. Bailey, and B. Campbell, Z. Phys. C
{\bf 61}, 613 (1994).

\bibitem{Langacker81}  P. Langacker, Phys. Rep. {\bf 72}, 185 (1981).

\bibitem{Georgi74}  H. Georgi, in {\it Particles and Fields} 1974, ed. C.
Carlson (AIP, NY, 1975), p. 575.

\bibitem{Fritzsch75} H. Fritzsch and P. Minkowski, Ann. Phys. {\bf 93}, 193
(1975).

\bibitem{Deshpande92} N. Deshpande, E. Keith, and P. Pal,
Phys. Rev. D {\bf 46}, 2261 (1992); {\bf 47}, 2892 (1993).

\bibitem{Eichten80} E. Eichten and K. Lane, Phys. Lett. {\bf 90B}, 125 (1980).

\bibitem{Dimopoulos80} S. Dimopoulos, S. Raby, and P. Sikivie, Nucl. Phys. {\bf
B176}, 449 (1980); S. Dimopoulos, S. Raby, and G. Kane, Nucl. Phys. {\bf
B182}, 77 (1981).

\bibitem{Lykken94} J. Lykken and S. Willenbrock, Phys. Rev. D {\bf 49},
4902 (1994), and references therein.

\bibitem{Appelquist93} T. Appelquist and J. Terning, YCTP-P21-93.

\bibitem{CDF} CDF Collaboration, F. Abe {\it et al.}, Phys. Rev. D {\bf 50},
2966 (1994).

\bibitem{Dimopoulos93} S. Dimopoulos, S. Raby, and G. Kane, Ref.
\cite{Dimopoulos80}.

\bibitem{Shanker206} O. Shanker, Nucl. Phys. {\bf B206}, 253 (1982).

\bibitem{Deshpande83} N. Deshpande and R. Johnson, Phy. Rev. D {\bf 27},
1193 (1983).

\bibitem{Arisaka93} K. Arisaka et al., Phys. Rev. Lett. {\bf 70}, 1049 (1993).

\bibitem{Donoghue92} J. Donoghue, E. Golowich, and B. Holstein, {\it Dynamics
of the Standard Model} (Cambridge University Press, Cambridge, 1992).

\bibitem{Weinberg77} S. Weinberg, Trans. N.~Y. Acad. Sci. {\bf 38}, 185 (1977).

\bibitem{Ukawa92} A. Ukawa, in {\it Lattice 92, Proceedings of the
International Symposium on Lattice Field Theory}, Amsterdam, 1992, edited
by J. Smit and P. van Baal, Nucl. Phys. B (Proc. Suppl.) {\bf 30}, 3 (1993).

\bibitem{CLEO93} CLEO Collaboration, R. Ammar {\it et al.}, Phys. Rev. D
{\bf 49}, 5701 (1994).

\bibitem{D0} D0 Collaboration, S. Abachi {\it et al.}, Phys. Rev. Lett.
{\bf 72}, 2138 (1994).

\bibitem{Shanker204} O. Shanker, Nucl. Phys. {\bf B204}, 375 (1982).

\bibitem{Marciano93} W. Marciano and A. Sirlin, Phys. Rev. Lett. {\bf 71},
3629 (1993).

\bibitem{Britton92} D. Britton {\it et al.}, Phys. Rev. Lett. {\bf 68},
3000 (1992).

\bibitem{Czapek93} C. Czapek {\it et al.}, Phys. Rev. Lett. {\bf 70}, 17
(1993).

\bibitem{Bartelt94} CLEO Collaboration, J. Bartelt {\it el al.}, CLNS-94-1287
(1994).

\bibitem{Cinabro92} CLEO Collaboration, D. Cinabro, {\it Proceedings of the
Fermilab Meeting, Division of Particles and Fields 1992}, edited by
C. Albright, P. Kasper, R. Raja, and J. Yoh (World Scientific, Singapore,
1993),
p.~843.

\bibitem{GW} T. Goldman and W. Wilson, Phys. Rev. D {\bf 15}, 709 (1977);
J. Bijnens, G. Ecker, and J. Gasser, Nucl. Phys. {\bf B396}, 81 (1993).

\bibitem{Heintze76} J. Heintze, Phys. Lett. {\bf 60B}, 302 (1976).

\end{thebibliography}
\end{document}